\documentclass[preprint,showpacs,pre,12pt]{revtex4-1}
\usepackage{amsfonts}
\usepackage{amssymb}
\usepackage{amsmath}
\usepackage{graphicx}
\usepackage{pgfplots}

\begin{document}
\title{Barrier billiard and random matrices}
\author{Eugene Bogomolny}
\affiliation{LPTMS,  CNRS, Univ. Paris-Sud, Universit\'e Paris-Saclay, 91405 Orsay, France}
\date{\today}
\begin{abstract}
The barrier billiard is the simplest example of pseudo-integrable models with 
 interesting and intricate classical and quantum properties. Using the Wiener-Hopf method it is demonstrated that quantum mechanics of a rectangular billiard with a barrier in the centre can be  reduced  to the investigation of a certain unitary matrix. Under heuristic assumptions this matrix is substituted by a special low-complexity random unitary matrix     
of independent  interest. The main results of the paper are (i) spectral statistics of such billiards is insensitive to the barrier height and (ii) it is well described by the semi-Poisson distributions. 
\end{abstract}

\maketitle

\section{Introduction}

An implicit idea  of quantum chaos studies is that quantum dynamics of even simple deterministic systems is so irregular and complex that the calculation of particular values of eigenenergies and eigenfunctions, though possible, leads to quasi-random quantities which may and have to be  substituted by  a statistical  description  of  such quantum problems. 

There are two big conjectures in quantum chaos:
\begin{itemize}
\item Local spectral statistics of generic quantum systems corresponding to classically   integrable systems is well described by the Poisson statistics of independent random variables \cite{berry_tabor}.
\item Local spectral statistics of generic quantum systems corresponding to classically chaotic systems is described by eigenvalue statistics of  standard ensembles of random matrices  depended only on system symmetry \cite{BGS}.   
\end{itemize}
Though these conjecture will, probably, never be proved in the full generality and there exist noticeable exceptions,  they form a cornerstone of quantum chaos and have been checked in  
enormous number of examples. 

Nevertheless, these conjectures do not cover all possible types of dynamical systems. For simplicity,  let us concentrate on  2-dimensional Hamiltonian models. Classically integrable systems are characterised by the condition that a typical trajectory belongs to a torus (i.e.,  a 2-dimensional surface of genus $1$). For classically chaotic models typical trajectories cover the whole 3-dimensional surface of constant energy. But there exist systems whose trajectories spread over 2-dimensional  surfaces of genus higher than $1$.  Such systems are neither integrable or chaotic and coined the name of pseudo-integrable models (see, e.g. \cite{richens_berry}). A characteristic example of such systems is a plane polygonal  billiard whose internal angles $\alpha_j$  are rational fractions of $\pi$: 
\begin{equation}
\alpha_j=\frac{m_j}{n_j} \pi
\end{equation}
with co-prime integers $m_j$ and $n_j$. It has been proved \cite{katok} that it this case classical trajectories belong to a surface of genus 
\begin{equation}
g=\frac{N}{2}\sum_{j} \frac{m_j-1}{n_{j}} 
\label{genus}
\end{equation}
where $N$ is the least common multiply of all denominators $n_j$. About classical dynamics of such billiards see, e.g., \cite{gutkin}, \cite{zorich} and references therein.  

The knowledge of quantum properties of pseudo-integrable billiards is fragmentary and includes  mainly  numerical calculations of statistical properties  of eigenenergies for billiards of simple shape: rhombus, right triangles, rectangular  billiard with a barrier, etc., \cite{cheon}-\cite{wiersig}. The only quantity accessible  analytically in certain models  is  the spectral compressibility $\chi$  which determines the growth of the variance of number of levels in an interval of length $L$ \cite{rigidity} 
\begin{equation}
\langle (N(E)-L)^2\rangle \underset{L\to\infty}{\longrightarrow} \chi L 
\end{equation}
where $N(L)$ is a number of levels in an interval $L$ normalised that its mean value equals $L$ and the averaging is taken over a small window of energies. The value of $\chi$ is of importance as for integrable models $\chi=1$ and for chaotic ones $\chi=0$ \cite{rigidity}. The calculation of the compressibility is done  by the  summation  over classical periodic orbits in the diagonal approximation \cite{rigidity}. For pseudo-integrable billiards the description of periodic orbits is known analytically for special class of billiards called the Veech billiards  \cite{veech, vorobets}, \cite{zorich}. In particular, for a right triangle with one angle $\pi/n$ in \cite{communications} has been proved that
\begin{equation}
\chi=\frac{n+\epsilon(n)}{3(n-2)}
\label{chi_n}
\end{equation}
where $\epsilon(n)=0,3,6$ for, respectively, odd $n$, even $n$ but $n\not \equiv 0 \mod 3$, $n\equiv 0 \mod 6$.

For the barrier billiard discussed below it has been shown (see \cite{wiersig} for the barrier height equals one-half of the billiard length, $h/a=1/2$,   and Appendix D of \cite{thesis} for an arbitrary height) that independently of the barrier height 
\begin{equation}
\chi=\frac{1}{2}. 
\label{chi_barrier}
\end{equation}
The fact that for these models $0<\chi<1$ is a clear-cut indication that spectral statistics of such billiards differ  from  the Poisson distribution typical for integrable models and from the random matrix statistics of chaotic systems. 

Numerically, it has been confirmed (cf.,  \cite{wiersig}, \cite{communications}) that the spectral statistics of the above billiards is special and is  characterised by following properties: 
\begin{itemize}
\item Level repulsion at small distances as for the standard random matrix ensembles.
\item Exponential decrease of the nearest-neighbour distributions as for the Poisson distribution.
\item Non-trivial value of the spectral compressibility (cf., \eqref{chi_n}, \eqref{chi_barrier}).
\item Multi-fractal dimensions of eigenfunctions \cite{wave_functions, superscars} .
\end{itemize}
This type of statistics has been first observed in the Anderson model at the point of the metal-insulator transition \cite{altshuler, schlovskii} and is called now an intermediate statistics. 

A canonical model of such statistics is the critical power-law random banded matrix  model  \cite{mirlin} (see also \cite{levitov, altshuler_levitov}) in which all matrix elements are independent Gaussian random variables with zero mean and the variances decreasing linearly from the main diagonal
\begin{equation}
\langle |H_{i,j}|^2\rangle=\left ( 1+\frac{|i-j|^2}{b^2}\right )^{-1}.
\end{equation}
This model has been thoroughly investigated (see, e.g., \cite{evers} and references therein) but its universality remains questionable. There exist several examples of matrices with intermediate type spectral statistics \cite{gerland_plasma}-\cite{integrable_ensembles} which clearly cannot be described by the above model.  In a sense, the critical power-law random banded matrix model is a minimal mathematical model which leads to intermediate statistics but it does not corresponds to a physical problem. 

The purpose of this work is twofold. First,  in Section~\ref{S_matrix_derivation}  it is demonstrated  that the investigation of the simplest pseudo-integrable model, the barrier billiard, can be reduced to the analysis of an  unitary $S$-matrix corresponding to the scattering on the barrier multiplied by certain phases related on the barrier height.   Using the Wiener-Hopf method, briefly reviewed in Appendix~\ref{wiener_hopf}, this matrix is calculated analytically. Second,  assuming that certain simple phases can be considered as random it is argued in Section~\ref{randomise_S_matrix}  that the exact $S$-matrix could be substituted by a random unitary matrix which belongs to a sub-class of low-complexity matrices with simple displacement structure \cite{complexity, displacement}. Using the same method as for  random Toeplitz and Hankel matrices \cite{toeplitz} it is shown in Section~\ref{properties_S_matrix} that local spectral statistics of the resulting random unitary matrix is well described by the semi-Posson distribution \cite{gerland_plasma} which agrees well with numerical calculations.  These results  imply that  eigenvalues of the barrier billiard are also statistically distributed by the same distribution. Section~\ref{conclusion} gives a brief summary of the obtained results.


\section{$S$-matrix approach to the quantisation of a barrier billiard}\label{S_matrix_derivation}

The 2-dimensional rectangular billiard is an archetype of integrable quantum models. Though its eigen-energies are trivial, e.g., for the Dirichlet boundary conditions $E_{m,n}=\pi^2 n^2/a^2+\pi^2 m^2/b^2$  where $a,\ b$ are side lengths and $m,\ n$ are positive integers, a rigorous treatment of its local spectral statistics is notoriously difficult due to the absence of explicitly random parameters. Only the two-point correlation function, ($R_2(s)=1$ in a convenient normalisation) is accessible to analytical calculations  \cite{marklof}. In physical literature it is conjectured that when $a^2/b^2$ is a 'good' irrational number (a Diophantine number?) then local spectral statistics of a rectangular billiard is well described by the Poisson statistics of independent random variables in accordance with the  existing numerics.  The proof or disproof of this conjecture seems to be beyond the known methods. 

\begin{figure}
\begin{minipage}{.49\linewidth}
\begin{center}
\includegraphics[width=.85\linewidth]{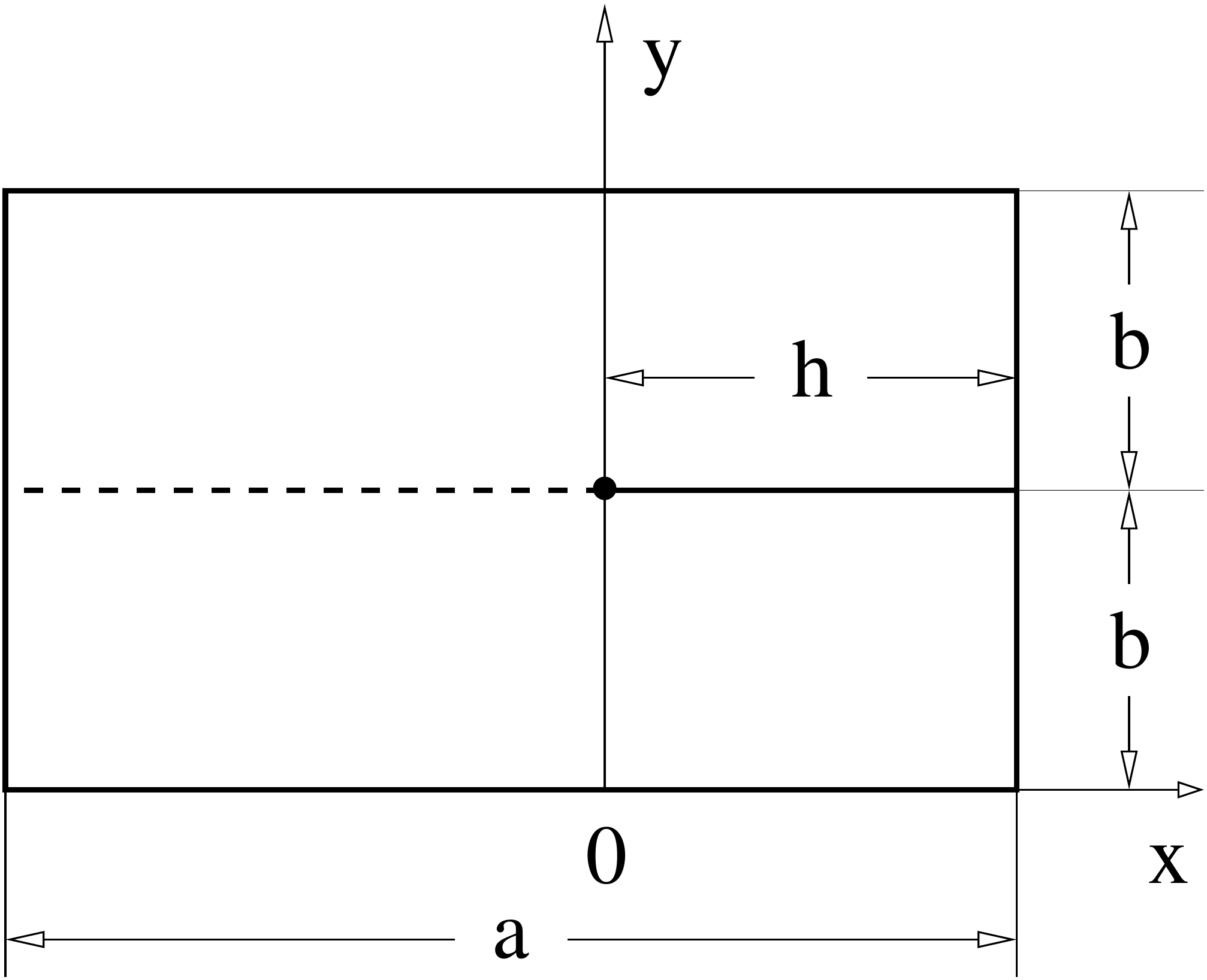}\\
(a)
\end{center}
\end{minipage}
\begin{minipage}{.49\linewidth}
\begin{center}
\includegraphics[width=.99\linewidth]{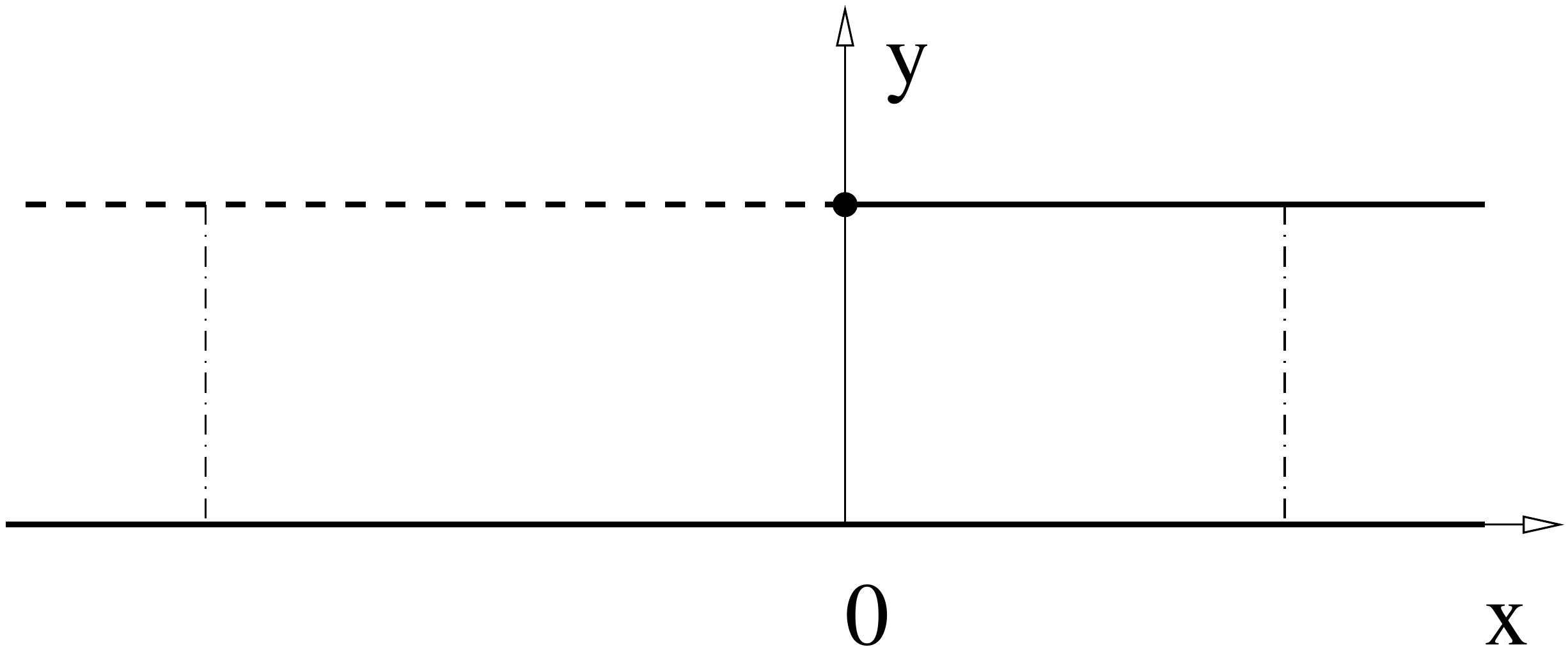}\\
(b)
\end{center}
\end{minipage}
\caption{(a) Barrier billiard. For clarity the tip of the barrier is indicated by a small circle. The ordinate axis passes through the barrier tip and the abscissa axis  passes through the lower side of the rectangle.  (b) The infinite slab with different boundary conditions along the the upper boundary.  Dashed-dotted lines indicate the sides of the initial rectangle. }
\label{barrier}
\end{figure}

The simplest pseudo-integrable model is the rectangular billiard with a barrier at the centre of a side  (see figure~\ref{barrier}(a)).  This polygon has 6 angles $\pi/2$ plus  angle $2\pi$ around the barrier tip. From \eqref{genus} it follows that it corresponds to a genus-two surface.

The quantisation of such billiard consists in finding the eigenvalues $E_{\alpha}$  and eigenfunctions $\Psi_{\alpha}(x,y)$ of the Helmholtz equation
\begin{equation}
\left ( \frac{\partial^2 }{\partial x^2} +\frac{\partial^2 }{\partial y^2}+E_{\alpha} \right )\Psi_{\alpha}(x,y)=0
\end{equation}
which obey the Dirichlet boundary conditions on all sides of the rectangle and on the barrier
\begin{equation}
\Psi_{\alpha}(x,y)|_{\mathrm{sides}}=0,\qquad \Psi_{\alpha}(x,y)|_{\mathrm{barrier}}=0.
\end{equation}
Due to the symmetry one set of solutions which equals zero at the whole line $y=b$ is evident
\begin{equation}
\Psi_{\alpha}(x,y)=\sin\Big ( \frac{\pi n}{a} (x-h)\Big ) \sin \Big ( \frac{\pi m}{b}y\Big) ,\qquad n,m=1,2, \ldots \ .
\end{equation}
We are interested in non-trivial solutions which are symmetric with respect to the inversion in the line passing through the barrier. In the coordinates as in figure~\ref{barrier}(a) it means that these solutions have to obey two sets of boundary  conditions 
\begin{eqnarray}
\Psi_{\alpha}(x,b)&=&0, \qquad 0< x < h,\nonumber\\
\frac{\partial }{\partial y}\Psi_{\alpha}(x,b)&=&0,\qquad h-a < x <0,\label{horisontal}\\
\Psi_{\alpha}(x,0)&=&0,\qquad h-a< x<h ,\nonumber
\end{eqnarray} 
and 
\begin{equation}
\Psi_{\alpha}(h,y)=0,\quad \Psi_{\alpha}(h-a,y)=0,\qquad 0<y<b.
\label{vertical}
\end{equation} 
No analytical solutions of the Helmholtz equation with such boundary conditions are known.

Let us disregard the vertical conditions \eqref{vertical} and find the scattering solutions of the infinite slab indicated in figure~\ref{barrier}(b). It implies that we are now looking for the solutions of the equation
\begin{equation}
\left ( \frac{\partial^2 }{\partial x^2} +\frac{\partial^2 }{\partial y^2}+k^2 \right )\Psi(x,y)=0
\end{equation} 
inside the slab such that at horizontal boundaries they obey the following conditions 
\begin{eqnarray}
\Psi(x,b)&=&0, \qquad 0< x < \infty,\nonumber\\
\frac{\partial }{\partial y}\Psi_{\alpha}(x,b)&=&0,\qquad -\infty < x <0,\label{horizontal_slab}\\
\Psi_{\alpha}(x,0)&=&0,\qquad -\infty< x<\infty .\nonumber
\end{eqnarray} 
As it is well known, to uniquely define such solutions one has to fix the behaviour on the infinity. 
 
The elementary solutions on negative and positive $x$ with fixed energy  have evidently the following forms   (the normalisation of plane waves to the unit current is used)
\begin{eqnarray}
\psi_{2m}^{(\pm)}(x,y)&=&\frac{e^{\pm ip_{2m} x}}{2\sqrt{b p_{2m}}}\sin\Big (\frac{\pi m}{b}y\Big ),\qquad x>0,
\label{positive_x}\\
\psi_{2m-1}^{(\pm)}(x,y)&=&\frac{e^{\pm ip_{ 2m-1} x}}{2\sqrt{b p_{2m-1}}}\sin\Big (\frac{\pi (2m-1)}{2b}y\Big ),\qquad x<0
\label{negative_x}
\end{eqnarray}
where 
\begin{equation}
p_m=\sqrt{k^2-\frac{\pi^2 m^2}{4b^2}}, \qquad m=1,2,\ldots,  \ .
\label{p_m_definition}
\end{equation}
There exit two sets of standard solutions determined by fixing the incoming plane waves. Any of  such solutions  can be expanded into corresponding series of elementary waves  \eqref{positive_x} and \eqref{negative_x}. 

For waves coming from the left one has the following expansion into reflected and transmitted waves
\begin{equation}
\Phi_{2n-1}^{(+)}(x,y)=\left \{ \begin{array}{cc} 
\phi_{2n-1}^{(+)}(x,y)+\sum_{m=1}^{\infty} S_{2n-1, 2m-1}\phi_{2m-1}^{(-)}(x,y), & x<0\\
\sum_{m=1}^{\infty} S_{2n-1,2m}\phi_{2m}^{(+)}(x,y),& x>0\end{array}\right . .
\label{left_waves}
\end{equation}
For waves coming from the right such expansion is 
\begin{equation}
\Phi_{2n}^{(-)}(x,y)=\left \{ \begin{array}{cc} 
\sum_{m=1}^{\infty} S_{2n,2m-1}\phi_{2m-1}^{(-)}(x,y),&x<0\\
\phi_{2n}^{(-)}(x,y)+\sum_{m=1}^{\infty} S_{2n,2m}\phi_{2m}^{(+)}(x,y),& x>0\end{array}\right . .
\label{right_waves} 
\end{equation}
The matrix $S_{mn}$ is the $S$-matrix for the scattering inside the slab. In Appendix~\ref{wiener_hopf} it is demonstrated that such matrix can be calculated analytically by the Wiener-Hopf method.   

By construction, functions $\Phi_{2n-1}^{(+)}(x,y)$ and $\Psi_{2n}^{(-)}(x,y)$ obey boundary conditions on horizontal boundaries \eqref{horizontal_slab}. To find  functions obeying the vertical conditions \eqref{vertical} let us form the linear combinations of these functions
\begin{equation}
\Psi_{\alpha}(x,y)=\sum_{n=1}^{\infty} a_{2n}\Psi_{2n}^{(-)}(x,y)+a_{2n-1}\Phi_{2n-1}^{(+)}(x,y). 
\end{equation}
Taking into account that functions \eqref{negative_x} and \eqref{positive_x} form complete set of functions at, respectively, negative and positive $x$,  the requirements \eqref{vertical}  signify  that for $m=1,2,\ldots, $
\begin{equation}
  a_{2m} +e^{2ip_{2m}h}\sum_{n=1}^{\infty} a_{n}S_{n,2m}, \qquad \mathrm{from} \; x=h,  
\end{equation} 
and 
\begin{equation}
a_{2m-1} +e^{2ip_{2m-1}(a-h)}\sum_{n=1}^{\infty} a_{n} S_{n,2m-1} \qquad \mathrm{from} \;x=h-a.
\end{equation}
Notice that the summation in these expressions are done over both  even and odd integers. 

Finally these equations can be rewritten for all $m$ as follows
\begin{equation}
a_{m}+\sum_{n=1}^{\infty} a_{n}B_{n,m}=0,\qquad m=1,2,\ldots, .
\end{equation}
where matrix $B_{n,m}$ differs from $S_{n,m}$ only by special phases
\begin{equation}
B_{n,m}=e^{i\phi_m} S_{n,m},\qquad \phi_{2m}=2p_{2m}h,\quad \phi_{2m-1}=2p_{2m-1}(a-h). 
\label{phases}
\end{equation}
The existence of such solutions determines the eigenvalue of $k$ from the quantisation condition
\begin{equation}
\det\big (\delta_{n,m}+B_{n,m}\big )=0. 
\end{equation}
Matrix $B$ contains the complete information about the quantisation of the barrier billiard. It constitutes of two parts: a specific  $S$-matrix for the scattering on a barrier and additional phases related with the position of the barrier. 


\section{Random matrix description of the barrier billiard}\label{randomise_S_matrix}

As it is shown in Appendix~\ref{wiener_hopf} \eqref{compact_S}, the scattering $S$-matrix is a complex symmetric matrix of the form
\begin{equation}
S_{n,m}=\frac{L_n\, L_m}{x_n+x_m}
\end{equation}  
where 
\begin{equation}
x_m=(-1)^{m+1}p_m
\label{x_m_billiard}
\end{equation}
 and vector   $L_n$ is  given by \eqref{explicit_L}.  

By construction,  matrix $B$ is 
\begin{equation}
B_{n,m}=e^{i\phi_n}\frac{L_n\, L_m}{x_n+x_m}
\end{equation}
where $\phi_n$ are defined in \eqref{phases}.  By conjugation this matrix can also  be transformed into symmetric shape but it is not necessary.   

For propagating modes with real $p_m$ matrix $S$ and, consequently matrix $B$ are unitary
\begin{equation}
SS^{\dag}=1,\qquad BB^{\dag}=1.
\end{equation}
Till now the calculations were exact.  Below we discuss 'natural' simplifications appeared in the semiclassical limit $k\to\infty$. The first remark is that   matrix $B$ includes the both, propagating (with real $p_m$) and evanescent (corresponding to imaginary $p_m$) modes. As evanescent modes in the semiclassical limit decay  exponentially quickly from the barrier tip one can neglect contributions of such modes provided that the tip is not very  close to the boundaries, $hk\gg 1$ and $(a-h)k\gg 1$. 

Then the $B$-matrix becomes  a finite dimensional   unitary matrix of (large) dimension 
 \begin{equation}
 N=[2k b/\pi]  
 \label{full_dimension} 
\end{equation}
which corresponds, in a sense, to an exact quantisation of a surface of section \cite{semiclassics, DS}. 

Eigenvalues of $N\times N$ unitary matrix $B\equiv B(k)$ with fixed parameter $k$ are of  the form $e^{i\epsilon_{\alpha}(k)}$ with real $\epsilon_{\alpha}(k)$.
Assume that $\epsilon_j(k)$ with fixed $k$ are ordered and restricted to an interval $[0,2\pi)$
\begin{equation}
0\leq \epsilon_1(k)<\epsilon_2(k)<\ldots<\epsilon_N(k)<2\pi. 
\end{equation}
True eigenenergies of the barrier billiard correspond to such values of $k$ for which one of eigenvalues of $B$ equals $-1$    
\begin{equation}
\epsilon_{\alpha}(k_{\alpha})=\pi .  
\label{epsilon_pi} 
\end{equation}
Below we cite  heuristic arguments from \cite{semiclassics, DS} that spectral statistics of eigenvalues of matrix $B$ and  of  barrier billiard eigenvalues are the same up to a rescaling. 
\begin{itemize}
\item The motion of eigenvalues of $B(k)$  when $k$ is changed from  $k=k_0$ to $k=k_0+\delta k$ with small $\delta k $ (such that $N(k)$ in \eqref{full_dimension} remains constant) can be approximated  as a sum of two terms, a smooth overall shift and a quasi-random contribution due to the scattering with other eigenvalues 
\begin{equation}
\epsilon_{\alpha}(k_0+\delta k)=\varepsilon_{\alpha} +\tau\, \delta k 
\end{equation}
\item Quantities $\varepsilon_{\alpha}$ are supposed to be so erratic function of $k$ that  their explicit form is irrelevant and they may be substituted by random numbers with certain correlation functions $R_n(x_1,\ldots, x_n)$  defined as the probability density that variable $\varepsilon_{\alpha}$ lies between $x_{\alpha}$ and $x_{\alpha}+dx_{\alpha}$.
\item The values of the true barrier billiard eigenmomenta $k_{\alpha}=k_0+\delta k_{\alpha}$  are  determined from \eqref{epsilon_pi} 
\begin{equation}
\delta k_{\alpha}=\gamma (\pi- \varepsilon_{\alpha} ),\qquad \gamma=\frac{1}{\tau} 
\end{equation}
\item The value of $\gamma$ can be estimated by comparison of the level density of unitary matrix eigenvalues, $d_B=N/(2\pi)$, and the level density of the barrier billiard in the momentum space,  $\bar{d}(k) =a b k/(2\pi)$,
\begin{equation}
\gamma d_B=\bar{d}(k),\qquad \gamma \approx  \frac{a}{4}. 
\end{equation}
\item If correlation functions $R_n(x_1,\ldots, x_n)$ are translation invariant, i.e., they depend only on the differences between eigenvalues, then  spectral statistics of barrier billiard eigenenergies is (up to a rescaling) the same spectral statistics of eigenvalues of matrix $B(k)$.   
\end{itemize}
Matrix  $B(k)$ has no explicit random parameters. As it is typical in quantum chaos 
pseudo-randomness of its eigenvalues and eigenfunctions  comes, supposedly,  from  erratic behaviour of its elements when parameter $k$ is changed.  This statement, though physically natural, is difficult to prove rigorously (if any). To get a well defined random matrix we assume that  in the semiclassical limit $N\to\infty$ deterministic exponential factors for propagating modes  $e^{i\phi_m}$ where $p_m$ as in \eqref{p_m_definition}   can be substituted  by  $e^{i\Phi_m}$ where $\Phi_m$ with $m=1,\ldots,N$  are independent random variables distributed uniformly between $0$ and $2\pi$.

Such assumption is also  not easy to prove.  It is similar to 'physical' statement that local statistics of rectangular billiard, and a posteriori of 'generic' integrable systems is well approximated by the Poisson statistics \cite{berry_tabor}. Nevertheless, the combination of the following facts: (i)  in the semiclassical limit $k\to\infty$  phases $\phi_m$ are large (except ones very close to the threshold of evanescent modes) and (ii) these phases are, in general, non-commensurable,  permit to conjecture that   quantities  $\phi_m$ mod $2\pi$ become pseudo-random (may be  after an averaging over a small window of $k$).  Though, in general, it may be true, there are proven counterexamples. In particular, the sequence $\sqrt{m}$  mod $1$ with $m=1,\ldots,N$  is uniformly distributed for large $N$ and its two-point correlation function agrees with the Poisson point process \cite{two_point_sqrt_n},  but  its nearest-neighbour distribution differs  from the Poisson expression \cite{sqrt_n}. 

\begin{figure}
\begin{center}
\includegraphics[width=.7\linewidth]{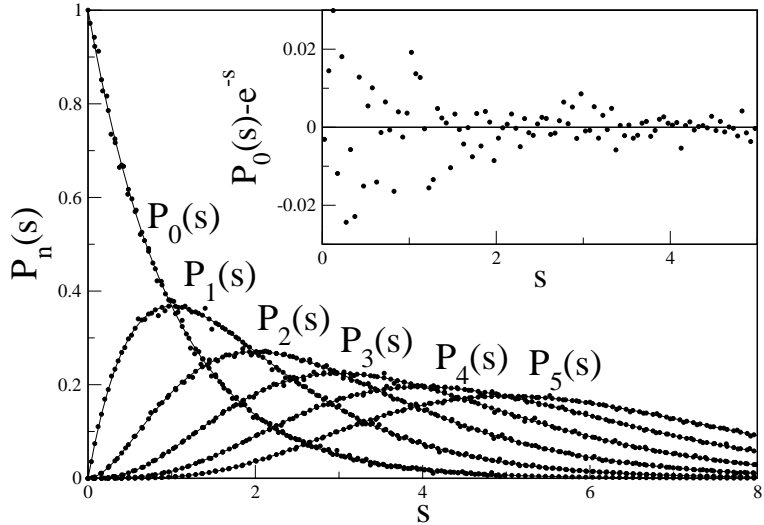}
\end{center}
\caption{The nearest-neighbour distributions with $n=0,\ldots,5$  computed numerically for pseudo-random phases  \eqref{scaled_phases} mod $2\pi$  with $N=10^5$, $\alpha=1$  and $\delta=1/2$ (black circles).   Solid lines are the Poisson predictions for these quantities \eqref{P_n_Poisson}. Insert: the difference between $P_0(s)$ and the Poisson formula: $P_{0}(s)= e^{-s}$. }
\label{P_n_random_phases}
\end{figure}

After a rescaling of $k$, the phases $\phi_m$ can be simplified as follows
\begin{equation}
\phi_m=\alpha \sqrt{(N+\delta)^2-m^2}, \qquad m=1,\ldots,N
\label{scaled_phases}
\end{equation}
with a constant $\alpha$ and $<\delta< 1$.  To check the validity of the above assumption for such phases numerical calculations  of $\phi_m$ mod $2\pi$ were performed. 
In figure~\ref{P_n_random_phases} the numerical  results for the nearest-neighbour distributions of these quantities are presented for $n=0,1,\ldots,5$ and $N=10^{5}$. In the calculations values $\delta=1/2$ and $\alpha=1$ were chosen but the results seems to be insensitive  to specific choices of these parameters. Solid lines in this figure indicate the well-known Poisson expressions for independent identically distributed uniform random variables
\begin{equation}
P_n(s)=\frac{s^n}{n!} e^{-s}. 
\label{P_n_Poisson}
\end{equation} 
It is clearly seen that the random phase  approximation works well for functions \eqref{scaled_phases}. To see better the accuracy of such approximation the difference between the numerical nearest-neighbour distribution $P_0(s)$ and the Poisson value $P_0(s)=e^{-s}$ is plotted in the Insert of this figure.

Taking  the  above arguments as granted allow us  to substitute  the deterministic unitary matrix $B$  by the  random unitary matrix 
\begin{equation}
B_{n,m}=\frac{e^{i\Phi_n}L_{n} L_{m}}{x_n+x_m},\qquad n,m=1,\ldots, N
\label{random_matrix}
\end{equation} 
where  $L_m$ by conjugation can be transformed into real quantities related with $x_j$ as follows \begin{equation}
L_m=\sqrt{2x_m\prod_{j\neq m} \frac{x_m+x_j}{x_m-x_j}},
\label{real_L}
\end{equation} 
$x_m$ are real quantities obeying the chain of inequalities (which is a consequence of the positivity of $L_m^2$)
\begin{equation}
x_1>-x_2>x_3>-x_4,\ldots, >0 , 
\label{inequalities}
\end{equation}
and $\Phi_m$ are independent random variables uniformly distributed between $0$ and $2\pi$. 

It is straightforward to check that any matrix as in \eqref{random_matrix} such that modulus $L_m$ is given by \eqref{real_L} is automatically unitary for arbitrary phases $\Phi_m$. 

All information about the barrier height  is contained in phases $\phi_m$ \eqref{phases}. After the replacement of these deterministic phases by random variables   this information is dislodged which means that spectral statistics of the barrier billiard in semiclassical limit  is independent on the barrier height. It concurs with the fact that the spectral compressibility \eqref{chi_barrier} is the same  for all barrier heights \cite{thesis} and with the results  \cite{wiersig} that numerically spectral statistics of the barrier billiard with $h/a=1/2$ and with  an irrational ratio $h/a$ look similar.  


\section{Properties of the main random matrix}\label{properties_S_matrix}

Matrix \eqref{random_matrix} belongs to the class of low-complexity matrices \cite{complexity} characterised by the following displacement operator \cite{displacement}  
\begin{equation}
\Delta_A(B)=A\, B+B\, A
\end{equation}
where matrix $A$ is a diagonal matrix $A_{i,j}=x_j\delta_{i,j}$.  From definition \eqref{random_matrix} it follows that 
\begin{equation}
\Delta_A(B)=e^{i\Phi_n}L_{n} L_{m}
\end{equation}
which implies  that the displacement operator of matrix $B$ is a rank-one matrix. According to a theorem proved in \cite{displacement}, principal matrix operations such as the matrix inversion and the calculation of  matrix eigenvalues  for matrices with finite displacement rank  can be performed in $\mathcal{O}(N^2)$ operations  to compare with $\mathcal{O}(N^3)$ operations needed for general matrices.  Here $N$ is the matrix dimension. 

It has been stressed in \cite{toeplitz} that random low-complexity matrices are good candidates for 
matrices with intermediate spectral statistics discussed in Introduction. The detailed investigation of statistical properties of matrix $B$ defined in \eqref{random_matrix} will be given elsewhere. Only main features of such matrix are discussed here. 

The local statistical properties of the eigenvalues spectrum are encoded in the nearest-neighbour distributions $P_n(s)$ which determine the probability densities that between two levels at a distance $s$ there exist exactly $n$ other levels.  The exact expressions for correlation functions of  matrices such as in \eqref{random_matrix} are unknown. To obtain simple  approximate Wigner-type formulas for these quantities we use the method developed  in \cite{toeplitz} for random Toeplitz and Hankel matrices. 

According to this method the nearest-neighbour distributions are well approximated by the gamma-distributions 
\begin{equation}
P_n(s)\approx a_n s^{\gamma_n} \exp\left (-b_n s\right ).
\label{gamma_dist}
\end{equation}  
If $\gamma_n$ is known, constants $a_n$ and $b_n$ are determined from the standard normalisation conditions
\begin{equation}
\int_0^{\infty} P_n(s) ds=1,\qquad \int_0^{\infty}s P_n(s) ds=n+1. 
\end{equation}
It has been argued in  \cite{toeplitz} that 
\begin{equation}
 \gamma_n=q_n-1
 \end{equation}
 where $q_n$ is the minimal number of  parameters  (the co-dimension) needed  to get $n+2$ eigenvalues of the considered matrix equal to each other. 

Matrix $B$ without random phases is also an unitary matrix
\begin{equation}
B^{(0)}_{n,m}=\frac{L_n L_m}{x_n+x_m},\qquad B^{(0)} B^{(0)\dag }=1.
\end{equation}
As this matrix is a real symmetric matrix, it implies that $B^{(0)\,2}=1$.
In other words, eigenvalues of matrix $B^{(0)}$ equal $\pm 1$. 

It is straightforward to prove that
\begin{equation}
\mathrm{Tr}\, B^{(0)}=\sum_{m=1}^N\frac{L_m^2}{2x_m}=\frac{1}{2}\left (1-(-1)^{N}\right ). 
\end{equation}
Therefore the minimum dimension matrix with $n+2$ eigenvalues equal 1 (and $n+1$ eigenvalues equal $-1$) is matrix $B^{(0)}$ of dimension $N_n=2n+3$. When $N_n$ non-zero random phases $\exp (i\Phi_m)$ are added  the degeneracy of eigenvalues is lifted. As an overall phase is unessential to us, the total number of independent (random) parameters is $q_n=N_n-1=2n+2$. In this way one comes to the prediction that  for matrix \eqref{random_matrix}  
\begin{equation}
\gamma_n=2n+1
\end{equation}  
which exactly corresponds to the semi-Poisson statistics discussed in \cite{gerland_plasma} for which 
\begin{equation}
P_n(s)=\frac{2^{2n+2} }{(2n+1)!} s^{2n+1} e^{-2s}.
\label{semi_poisson} 
\end{equation}
Besides random phases matrix $B$ depend on coordinates $x_m$. In principle, for the barrier billiard these variables are related with the momenta as indicated in \eqref{x_m_billiard}.  As this matrix is independent on the over-all scale of $x_m$ such `natural`  $x_m$ can conveniently be expressed as follows 
\begin{equation}
x_m=(-1)^{m+1} \sqrt{ (N+\delta)^2-m^2},\qquad m=1,\ldots,N
\label{x_momenta} 
\end{equation}
with  $0<\delta  <1$.
 
Nevertheless, the above conclusion that spectral statistics of matrix $B$ should be well described by the simple  semi-Poisson distribution \eqref{semi_poisson} is  valid for any sequence of $x_m$ (but obeying \eqref{inequalities}) which suggests that spectral statistics of  this matrix is only weekly depended  of the choice of coordinates $x_m$. 

To check these predictions numerical calculations of the nearest-neighbour distributions for matrix $B$ were performed for 3 different choices of $x_m$. The first corresponds to \eqref{x_momenta}, the second is the linear $x_m$
\begin{equation}
x_m=(-1)^{m+1}(N+1-m),\qquad m=1,\ldots, N
\label{x_linear}
\end{equation}
and for the third one $|x_m|$  are  chosen independently and uniformly  between $0$ and $N$,  then arranged to obey \eqref{inequalities}),  and  remained fixed for different realisations of  random phases.

\begin{figure}
\begin{center}
\includegraphics[width=.7\linewidth]{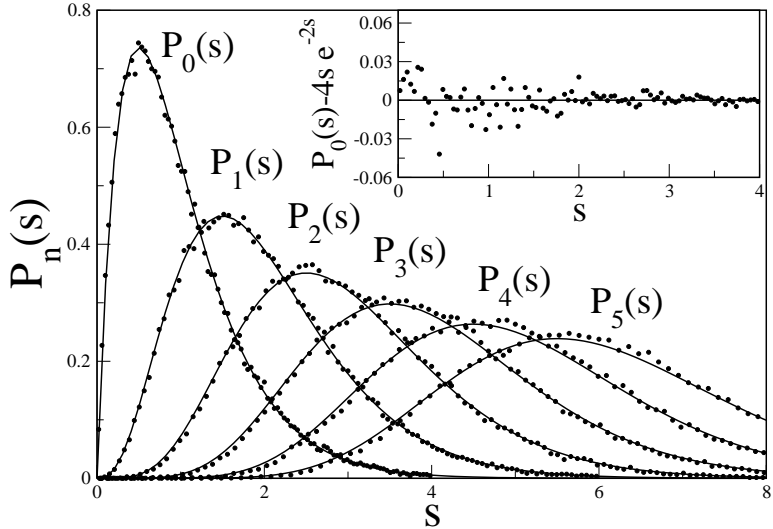}
\end{center}
\caption{The nearest-neighbour distributions with  $n=0,\ldots,5$ for $x_m$ as in \eqref{x_momenta} (black circles).   Solid lines are the semi-Poisson predictions for these quantities \eqref{semi_poisson}. Insert: the difference between $P_0(s)$ and the semi-Poisson formula: $P_{0}(s)=4s e^{-2s}$. }
\label{P_n_momentum}
\end{figure}

\begin{figure}
\begin{center}
\includegraphics[width=.7\linewidth]{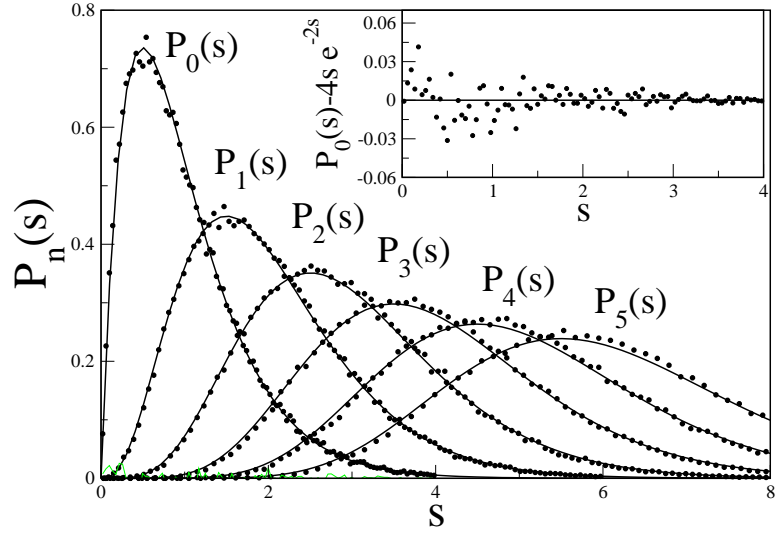}
\end{center}
\caption{The same as in figure~\ref{P_n_momentum} but for linear $x_m$ as in \eqref{x_linear}.  }
\label{P_n_linear}
\end{figure}

\begin{figure}
\begin{center}
\includegraphics[width=.7\linewidth]{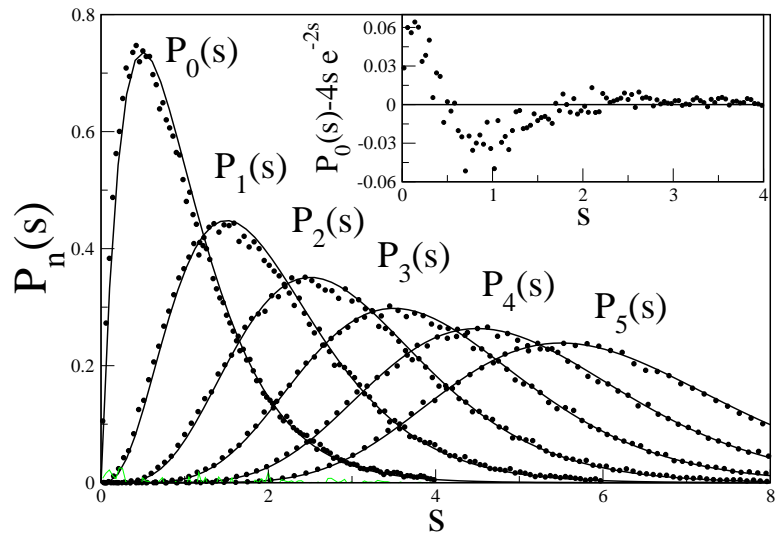}
\end{center}
\caption{The same as in figure~\ref{P_n_momentum} but for random $x_m$. }
\label{P_n_random}
\end{figure}

The results of these calculations are presented in figures~\ref{P_n_momentum}-\ref{P_n_random}. The calculations were done for matrices of dimension $N=1000$ averaged over  $100$ realisations
 of random phases $\Phi_m$ chosen independently and uniformly between $0$ and $2\pi$. 
 
 To see clearly the differences between the three different choices of variables $x_m$ the  corresponding data  are indicated at different figures: figure~\ref{P_n_momentum}  shows the data  when $x_m$ are chosen as in \eqref{x_momenta} with $\delta=.5$ (results seems to be insensitive to $\delta>0$), figure~\ref{P_n_linear} displays the data for linear choice of $x_m$ as in \eqref{x_linear}, and figure~\ref{P_n_random}  exhibits the results for random choice of $x_m$. In each figures small circles indicate numerical results for  $6$ nearest-neighbour distributions $P_n(s)$ with $n=0,1,\ldots,5$. The solid lines are the semi-Poisson predictions \eqref{semi_poisson}.  The differences between the nearest-neighbour distribution $P_0(s)$ and the semi-Poisson formula $P_0(s)=4s e^{-2s}$ are presented in the Inserts  of these figures.   
 
The figures clearly demonstrate that simple  approximate semi-Poisson formulas \eqref{semi_poisson} agree quite well with numerical results for different local correlation functions of random matrix $B$. As expected, the results for different choices of variables $x_m$ are close to each others but the data for random $x_m$ seems to have larger (and more regular) deviations from the semi-Poisson predictions.   
 
\section{Conclusion}\label{conclusion}

 The main result of the paper is the derivation of a random matrix associated with the pseudo-integrable barrier billiard.  It is  demonstrated  that  the quantisation of the barrier billiard can conveniently be performed by a two-steps procedure.  First, two boundaries of the  billiard are removed and the problem is reduced to the scattering inside of an infinite  slab with  different boundary conditions (the Dirichet and the Newman ones) along one boundary. The exact solution for this configuration is done by the Wiener-Hopf  method. Second, an eigenfunction  of  the closed billiard is represented as a linear combination of obtained scattering waves and the requirement that such eigenfunction obeys the correct boundary conditions on previously removed boundaries leads to the quantisation condition that a certain unitary matrix has an eigenvalue equals $-1$. 
 
 The resulting matrix differs from the $S$-matrix for the scattering inside the infinite slab only by certain phases related with the position of the barrier.  In principle, it could serve  for numerical calculations of quantum properties of the barrier billiard. But in the context of the paper,  its principal  importance is due to  the fact that under 'physical' assumptions the exact matrix  can be substituted by a random unitary matrix of a special form. An immediate consequence of such  replacement is that spectral statistics of the considered barrier billiard is independent on the barrier height. 
 
  It seems that it is the first time that a random matrix has been extracted from the exact quantum-mechanical  description of a pseudo-integrable model. The resulting random unitary matrix belongs to the so-called low-complexity matrices with interesting statistical properties and is of  independent interest.  It is demonstrated that local spectral statistics of this matrix are   well  approximated by the so-called semi-Poisson distribution in  accordance with numerical calculations of the nearest-neighbour distributions. As discussed in the text, it implies that spectral statistics of the barrier billiard has to be also close to the semi-Poisson statistics. 


\appendix
\section{Construction of the $S$-matrix for the slab by the Wiener-Hopf method}\label{wiener_hopf}

The purpose of this Appendix is to calculate explicitly the $S$-matrix for the scattering inside the slab  indicated in figure~\ref{barrier}(b). Due to the special geometry  of the slab  the Wiener-Hopf method \cite{noble} seems to be ideally suited for this purpose. Though this old method is well known (see e.g., \cite{noble}), for completeness,  the main steps of the solution of this problem are briefly indicated below.    

Consider the incident plane wave $\psi_{2n-1}^{(+)}(x,y)$ as in \eqref{negative_x}  entering the slab  from the left in figure~\ref{barrier}(b).  The total field inside the slab is the sum of the incident field and the reflected field $\psi(x,y)$ 
\begin{equation}
\Psi(x,y)=e^{ip_{2n-1} x}\sin \Big (\frac{\pi (2n-1)}{2b} y\Big)+\psi(x,y).
\end{equation}
As it is inherent in the Wiener-Hopf method \cite{noble} one assumes that the momentum $k$ has a small positive imaginary part so  $\mathrm{Im}\, p_m>0$ and the reflected field is determined by requirement that $\psi(x,y)\underset{|x|\to\infty}{\longrightarrow} 0$. 
By construction  the total field has to obey boundary conditions indicated in \eqref{horizontal_slab}. 

To obtain the Wiener-Hopf equation we follow closely the method of \cite{noble}. Define
\begin{equation}
\Phi_+(\alpha,y)=\int_0^{\infty}\psi(x,y)e^{i\alpha x} dx,\qquad \Phi_-(\alpha,y)=\int_{-\infty}^0\psi(x,y)e^{i\alpha x} dx.
\end{equation}
Here $\alpha$ is a complex variable such that
\begin{equation} 
-\mathrm{Im}\, k< \mathrm{Im} \, \alpha< \mathrm{Im}\, k.
\label{common_part}
\end{equation}
From boundary conditions \eqref{horizontal_slab} one gets the boundary values of  $\Phi_{\pm}(\alpha,b)\equiv  \Phi_{\pm}(\alpha)$ 
\begin{equation}
\Phi_+(\alpha)+\int_0^{\infty} e^{i (p_{2n-1}+\alpha)x}dx \sin\Big ( \frac{\pi (2n-1)}{2}\Big )=0,\qquad 
\Phi_+(\alpha)=\frac{i(-1)^n}{\alpha+p_{2n-1}}
\end{equation} 
and
\begin{equation}
\frac{\partial}{\partial y}\Phi_{-}(\alpha)=0.
\end{equation}
It is plain that $\Phi(\alpha,y)=\Phi_+(\alpha,y)+\Phi_-(\alpha,y)$ obeys the equation
\begin{equation}
\left ( \frac{\partial^2}{\partial y^2}+q^2(\alpha)\right )\Phi(\alpha,y)=0,\qquad q(\alpha)=\sqrt{k^2-\alpha^2}.
\end{equation}
Its solution equal zero at $y=0$ is  
\begin{equation}
\Phi(\alpha,y)=A(\alpha)\sin(q(\alpha) y) 
\end{equation}
where $A(\alpha)$ is a certain function.

Evaluating this expression at $y=b$ gets two equations
\begin{eqnarray}
\Phi_{-}(\alpha)+\frac{i(-1)^n}{\alpha+p_{2n-1}}&=&A(\alpha)\sin (q(\alpha) b)\label{A_alpha}\\
\frac{\partial}{\partial y}\Phi_{+}(\alpha)&=&q A(\alpha) \cos (q(\alpha) b)
\nonumber
\end{eqnarray}
Removing $A(\alpha)$ from these equations leads to the standard Wiener-Hopf equation
\begin{equation}
\Phi_{-}(\alpha)+\frac{i(-1)^n}{\alpha+p_{2n-1}}=b K(\alpha) \frac{\partial}{\partial y} \Phi_{+}(\alpha), \qquad K(\alpha)=\frac{\tan(q(\alpha)b)}{q(\alpha)b}.
\label{main_equation}
\end{equation}
The principal step in the Wiener-Hopf method is the factorisation of  $K(\alpha)$ 
\begin{equation}
K(\alpha)=K_{+}(\alpha)K_{-}(\alpha) 
\end{equation}
where $K_{+}(\alpha)$ has no zero and singularities  in the upper half-plane $\mathrm{Im}\, \alpha>-\mathrm{Im}\, k$ and $K_{-}(\alpha)$ is free of zero and singularities in the lower half-plane  $\mathrm{Im}\, \alpha< \mathrm{Im}\, k$. 

Using well known formulas
\begin{equation}
\sin x=x\prod_{n=1}^{\infty}\Big (1-\frac{x^2}{\pi^2 n^2}\Big ),\qquad \cos x=\prod_{n=1}^{\infty}\Big (1-\frac{x^2}{\pi^2 (n-1/2)^2}\Big )
\end{equation}
it is plain that 
\begin{equation}
K_{+}(\alpha)=\prod_{n=1}^{\infty} \frac{\sqrt{k^2b_n^2-1}+\alpha b_n}{\sqrt{k^2b_{n-1/2}^2-1}+\alpha b_{n-1/2}},
\qquad K_{-}(\alpha)=K_{+}(-\alpha) .
\label{K}
\end{equation}
Here
\begin{equation}
 b_n=\frac{b}{\pi n},\qquad b_{n-1/2}=\frac{b}{\pi (n-1/2)}.
\end{equation}
Divided \eqref{main_equation} by $K_{-}(\alpha)$ and separating the pole at $\alpha=-p_{2n-1}$ one obtains
 \begin{eqnarray}
& & \frac{\Phi_{-}(\alpha)}{K_{-}(\alpha)}+\frac{i(-1)^n}{(\alpha+p_{2n-1})}\left (\frac{1}{K_{-}(\alpha) }-\frac{1}{K_{-}(-p_{2n-1})}\right ) =\nonumber\\
& &
b K_{+}(\alpha) \frac{\partial}{\partial y} \Phi_{+}(\alpha)-\frac{i(-1)^n}{(\alpha+p_{2n-1})K_{-}(-p_{2n-1})}.
\end{eqnarray}
The left-hand side of this equation is free of  singularities in the lower  half-plane of $\alpha$ and the right-hand side is  regular in the upper half-plane. These half-planes have a common part \eqref{common_part}, thus the both sides have to be analytic in the whole plane of complex variable 
$\alpha$, i.e., equal to a certain polynomial. From boundary conditions it follows that this polynomial is zero.   Therefore  
\begin{equation}
	\frac{\Phi_{-}(\alpha)}{K_{-}(\alpha)}+\frac{i(-1)^n}{(\alpha+p_{2n-1})}\Big (\frac{1}{ K_{-}(\alpha) }-\frac{1}{ K_{-}(-p_{2n-1}) }\Big ) =0
\end{equation}
and 
\begin{equation}
b K_{+}(\alpha)  \frac{\partial}{\partial y}b \Phi_{+}(\alpha)-\frac{i(-1)^n}{(\alpha+p_{2n-1})K_{-}(-p_{2n-1}) }=0. 
\end{equation}
From \eqref{A_alpha} it follows that 
\begin{eqnarray}
A(\alpha)&=&\frac{i(-1)^n}{\sin(qb) \,(\alpha+p_{2n-1})}\frac{K_{-}(\alpha)}{K_{-}(-p_{2n-1})}\nonumber\\
&=& 
\frac{i(-1)^n}{qb \cos(qb) \,(\alpha+p_{2n-1})}\frac{1}{K_{+}(\alpha) K_{-}(-p_{2n-1})}
\label{A_left}
\end{eqnarray}
The first expression is convenient for $x>0$ and the second one for $x<0$.

The knowledge of this function permits to calculate the reflected field by the inverse Fourier transform
\begin{equation}
\psi(x,y)=\frac{1}{2\pi} \int_{-\infty }^{\infty } A(\alpha)\sin (q(\alpha) y )e^{-i\alpha x} d\alpha
\end{equation} 
For $x>0$ one can shift the integration contour into the lower half-plane of $\alpha$.  As $K_{-}(\alpha)$ has no singularity here the poles come only from zeros of $\sin\big (q(\alpha)b\big)$  plus a pole at  $\alpha=-p_{2n-1}$. One has
\begin{equation}
\sin(qb)=0\longrightarrow q=\frac{\pi m}{b}\longrightarrow \alpha_m=-p_{2m}=-\sqrt{k^2-\frac{\pi^2 m^2}{b^2}} ,\quad m=1,2,\ldots, .
\end{equation}
The residue at this point is 
\begin{equation}
\frac{\partial}{\partial \alpha} \sin (qb)\Big |_{\alpha=-p_{2m}}=b\frac{\partial q(\alpha) }{\partial \alpha} \cos(q(\alpha) b)\Big |_{\alpha=-p_{2m}}= \frac{b^2 (-1)^m p_{2m}}{\pi m}
\end{equation}
The contribution from the pole at $\alpha=-p_{2n-1}$ cancels the incident field and in the end one gets that for $x>0$ the total transmitted field is as in \eqref{left_waves} with 
\begin{equation}
 S_{2n-1,2m}=\frac{(-1)^{m+n} \pi m}{b^2 \sqrt{p_{2m-1} p_{2m}} (p_{2n-1}-p_{2m})} \frac{K_{+}(p_{2m})}{K_{+}(p_{2n-1})}. 
 \label{S_2n-1_2m}
\end{equation}
For $x<0$ one can shift the integration contour in the upper half-plane. The only singularities of the second expression  in \eqref{A_left}  are poles at points where $\cos\big (q(\alpha)b\big )=0$ or 
\begin{equation}
q=\frac{\pi}{b}(m-1/2), \qquad \alpha=p_{2m-1}, \qquad m=1,2,\ldots, 
\end{equation}
and 
\begin{equation}
\frac{\partial}{\partial \alpha} \cos (qb)\Big |_{\alpha=p_{2m-1}}=-b\frac{\partial q(\alpha) }{\partial \alpha} \sin (q(\alpha) b)\Big |_{\alpha=p_{2m-1}}= -\frac{b^2 (-1)^m p_{2m-1}}{\pi (m-1/2) }.
\end{equation}
Combining all terms together one concludes that the reflected field has the form as in \eqref{left_waves} with 
\begin{equation}
S_{2n-1,2m-1}=\frac{(-1)^{m+n} }{b^2\sqrt{p_{2n-1} p_{2m-1}} (p_{2n-1}+p_{2m-1})K_{+}(p_{2m-1}) K_{+}(p_{2n-1})}. 
\end{equation}
In these expressions  the relation $K_{-}(-\alpha)=K_{+}(\alpha)$ was used.

Exactly the same method can be used to find the scattering field  for the incoming wave from $+\infty$ \eqref{right_waves} and the corresponding coefficients are 
\begin{equation}
 S_{2n,2m-1}=\frac{(-1)^{n+m}\pi n }{b^2\sqrt{p_{2n} p_{2m-1}} (p_{2m-1}-p_{2n})}
\frac{K_{+}(p_{2n})}{K_{+}(p_{2m-1})}
\end{equation}
and 
\begin{equation}
S_{2n, 2m}=-\frac{(-1)^{n+m}\pi^2 m n }{b^2\sqrt{p_{2n} p_{2m}} (p_{2n}+p_{2m})}
K_{+}(p_{2n})K_{+}(p_{2m}).
\end{equation}
The above expressions for the $S$-matrix can conveniently be  rewritten in the following compact form 
\begin{equation}
S_{n,m}=\frac{L_n L_m}{x_n+x_m}
\label{compact_S}
\end{equation}
where 
\begin{equation}
x_{2m-1}=p_{2m-1},\qquad x_{2m}=-p_{2m}
\label{x_m}
\end{equation}
and 
\begin{equation}
L_{2n-1}=\frac{(-1)^n}{b \sqrt{p_{2n-1}} K_{+}(p_{2n-1})},\qquad 
L_{2n}=\frac{(-1)^n \pi n \, K_{+}(p_{2n})}{b\sqrt{p_{2n}}}.  
\label{explicit_L}
\end{equation}
In general, there exist two types of waves, propagating and evanescent corresponding, respectively,  to real and imaginary values of momenta, $p_m=\sqrt{k^2-\pi^2m^2/4b^2}$.  There are $N_e$ propagating modes with even $m$ and $N_o$ with odd $m$
\begin{equation}
N_e=\left [ \frac{kb}{\pi} \right ],\qquad N_o=\left [ \frac{kb}{\pi}+\frac{1}{2} \right ]. 
\label{number_modes}
\end{equation}
The modulus of $L_m$ is determined by propagating modes. On has  an important relation
\begin{equation}
\left |L_m\right |^2=2x_m\prod_{n\neq m} \frac{x_m+x_n}{x_m-x_n}.
\label{modulus_L}
\end{equation} 
Indeed, from \eqref{K}  by separating propagating and evanescent modes it follows that
\begin{equation}
K_{+}(p_{2m} )=
\frac{\prod_{n=1}^{N_e} (p_{2m}+p_{2n})}{\prod_{n=1}^{N_o} (p_{2m}+p_{2n-1})}W_{2m}, \qquad 
W_{2m}=\frac{\prod_{n>N_e} (p_{2m}+p_{2n})}{\prod_{n>N_o} (p_{2m}+p_{2n-1})}\prod_{n}\left (1-\frac{1}{2n}\right ). 
\end{equation}
In $W_{2m}$ momenta $p_n$ are imaginary $p_n=i\sqrt{\pi^2 n^2/4b^2-k^2}$. Therefore 
\begin{equation}
\left |W_{2m}\right |^2=\Big (\frac{\pi}{b}\Big )^{N_e-N_o}\frac{\prod_{n=N_e+1}^{\infty} (n^2 -m^2)}{\prod_{n=N_o+1}^{\infty}((n-1/2)^2-m^2)}
\prod_{n=1}^{\infty}(1-1/(2n))^2. 
\end{equation}
This expression can be rewritten as follows
\begin{equation}
\left |W_{2m} \right |^2=  \frac{\prod_{ n \neq m}^{\infty} (n^2 -m^2)}{ \prod_{n=1}^{\infty}((n-1/2)^2-m^2)(1-1/(2n))^{-2}}\, 
\Big (\frac{b^2}{\pi^2}\Big )  \frac{\prod_{n=1}^{N_o}(p_{2m}^2-p_{2n-1}^2)}{\prod_{ n\neq m}^{N_e}(p_{2m}^2-p_{2n}^2)}
\end{equation}
The first product is equal  
\begin{equation}
\lim_{x\to m} \frac{1}{m^2-x^2} \prod_{n=1}^{\infty}\frac{1-x^2/n^2}{1-x^2/(n-1/2)^2}=\lim_{x\to m} \frac{1}{m^2-x^2}\frac{\tan \pi x}{\pi x}=-\frac{1}{2m^2}.
\end{equation}
Using the definition \eqref{x_m} one gets \eqref{modulus_L} for even indices. Similar arguments prove \eqref{modulus_L} for odd indices.



\begin{thebibliography}{99}

\bibitem{berry_tabor} M. V. Berry and M. Tabor, \textit{Level clustering in the regular spectrum}, Proc. R. Soc. Lond. A \textbf{356},  375, (1977).

\bibitem{BGS} O. Bohigas, M. J. Giannoni, and C. Schmit, \textit{Characterization of chaotic quantum spectra and universality of level fluctuation laws},  Phys. Rev. Lett. \textbf{52}, 1 (1984).

\bibitem{richens_berry} P.J. Richens and M.V. Berry, \textit{Pseudointegrable systems in classical and quantum mechanics}, Physica D: Nonlinear Phenomena \textbf{2}, 495 (1981). 

\bibitem{katok} A.N. Zemlyakov and A.B. Katok, \textit{Topological transitivity in billiards in polygons}, Math. Notes \textbf{18}, 760 (1975).

\bibitem{gutkin} E. Gutkin, \textit{Billiards in polygons}, Physica D \textbf{19}, 311 (1986); E. Gutkin,  \textit{Billiards in polygons:survey of recent results}, J. Stat. Phys.\textbf{83}, 7 (1996).

\bibitem{zorich}  A. Zorich, \textit{Flat surfaces, On random matrices, zeta functions and dynamical systems}, Frontiers in Number Theory, Physics and Geometry, \textbf{1} (P. Cartier, B. Julia, P. Moussa, and P. Vanhove, eds.), Springer-Verlag, Berlin,  439 (2006).

\bibitem{cheon} T. Cheon and T. D. Cohen, \textit{Quantum level statistics of pseudointegrable billiards}, Phys. Rev. Lett. \textbf{62}, 2769 (1989).

\bibitem{shudo} A. Shudo and Y.  Shimizu, \textit{Extensive numerical study of spectral statistics for rational and irrational polygonal billiards}, Phys. Rev. E \textbf{47}, 54 (1993).

\bibitem{seba}  A. Shudo, Y.  Shimizu, Petr \u{S}eba, J. Stein, H.-J. St\"{o}ckmann, and 
K.  Zyczkowski, \textit{Statistical properties of spectra of pseudointegrable systems}, 
Phys. Rev. E \textbf{49}, 3748 (1994). 

\bibitem{schachner}   H. C. Schachner,  G. M. Obermair, \textit{Quantum billiards in the shape of right triangles} Z. Physik B - Condensed Matter \textbf{95}, 113 (1994). 

\bibitem{gerland} E. Bogomolny, U. Gerland, and C. Schmit, \textit{Models of intermediate spectral statistics}, Phys. Rev.E \textbf{59}, R1315 (1999).

\bibitem{jain} B. Gr\'{e}maud and S. R. Jain, \textit{Spacing distributions for rhombus billiards}, 
 J. Phys. A: Math. Gen. \textbf{31},  L637 (1998).
 
 \bibitem{communications} E. Bogomolny, O. Giraud, and C. Schmit, \textit{Periodic orbits contribution to the 2-point correlation form factor for pseudo-integrable systems}, Commun. Math. Phys. \textbf{222}, 327 (2001).
 
 \bibitem{gorin} T. Gorin, \textit{Generic spectral properties of right triangle billiards}, J. Phys. A: Math. Gen. \textbf{34}, 8281 (2001).

\bibitem{wiersig} Jan Wiersig, \textit{Spectral properties of quantized barrier billiards},
Phys. Rev. E \textbf{65}, 046217 (2002). 

\bibitem{rigidity} M. V. Berry, \textit{Semiclassical theory of spectral rigidity} Proc. Roy. Soc.
A \textbf{400}, 229 (1985).

\bibitem{veech} W. A.  Veech, \textit{ Teichm\"{u}ller curves in moduli space, Eisenstein series and an application to triangular billiards},  Invent. Math. \textbf{97}, 553 (1989).

\bibitem{vorobets}  Y. B. Vorobets, \textit{Planar structures and billiards in rational polygons:
the Veech alternative} Russian Math. Surveys \textbf{51}, 779  (1996). 

\bibitem{thesis} O. Giraud, \textit{Spectral statistics of diffractive systems}, PhD thesis (2002). 

\bibitem{wave_functions} E.Bogomolny and C. Schmit, \textit{Structure of wave functions of pseudo-integrable billiards}, Phys. Rev. Lett. \textbf{92}, 244102 (2004).

\bibitem{superscars} E. Bogomolny, \textit{Formation of superscar waves in plane polygonal
billiards},  J. Phys. Commun. \textbf{5},  055010 (2021).

\bibitem{altshuler} B. L. Altshuler, I. Kh. Zharekeshev, S. A. Kotochigava, \textit{Repulsion between levels and the metal-insulator transition}, Sov. Phys. JETP \textbf{67}, 625 (1988).

\bibitem{schlovskii} B.I. Shklovskii, B. Shapiro, B.R. Sears, P. Lambrianides, and H.B. Shore, \textit{Statistics of spectra ofdisordered systems near the metal-insulator transition}, Phys. Rev. B \textbf{47}, 11487 (1993).

\bibitem{mirlin}  A. D. Mirlin, Y. V. Fyodorov, F.-M. Dittes, J. Quezada, and T. H. Seligman, \textit{Transition from localized to extended eigenstates in the ensemble of power-law random banded matrices}, Phys. Rev. E \textbf{54}, 3221 (1996).

\bibitem{levitov} L. S. Levitov, \textit{Localization-delocalization transition for one-dimensional alloy potentials},  EPL  \textbf{7},   343 (1988).

\bibitem{altshuler_levitov}  B. L. Altshuler and S. Levitov, \textit{ Weak chaos in a quantum Kepler problem}, Phys. Rep. \textbf{288}, 487 (1997). 

\bibitem{evers} F. Evers and A. D. Mirlin, \textit{Anderson transitions}, 
Rev. Mod. Phys. \textbf{80}, 1355 (2008).

\bibitem{gerland_plasma} E. Bogomolny, U. Gerland, and  C. Schmit, \textit{Short-range plasma model for intermediate spectral statistics},   Eur. Phys. J. B \textbf{19}, 121 (2001).


\bibitem{interval} E. Bogomolny and C. Schmit, \textit{Spectral statistics of a quantum interval-exchange map}, Phys. Rev. Lett. \textbf{93}, 254102 (2004).


\bibitem{lax} E. Bogomolny, O. Giraud, and C. Schmit, \textit{Random matrix ensembles associated with Lax matrices},  Phys. Rev. Lett. \textbf{103}, 054103 (2009).

\bibitem{integrable_ensembles} E. Bogomolny, O. Giraud, and C. Schmit, \textit{Integrable random matrix ensembles}, Nonlinearity \textbf{24}, 3179 (2011). 


\bibitem{complexity} V. Y. Pan, Z. Q. Chen, and A. Zheng, \textit{The complexity of the algebraic eigenproblem}, STOC '99, Proc. of the thirty-first annual ACM symposium on theory of computing, 507 (1999).

\bibitem{displacement} T. Kailath, S.-Y. Kung, and M. Morf, \textit{Displacement ranks of matrices and linear equations}, J. Math. Anal. Applic. \textbf{68}, 395 (1979).

\bibitem{toeplitz} E. Bogomolny and O. Giraud, \textit{Statistical properties of structured random matrices},  Phys. Rev. E \textbf{103}, 042213 (2021). 

\bibitem{marklof} J. Marklof, \textit{Spectral form factors of rectangle billiards},  Comm. Math. Phys. \textbf{199}, 169 (1998).

 
\bibitem{semiclassics} E. Bogomolny, \textit{Semiclassical quantization of multidimensional systems},  Nonlinearity \textbf{5}, 805, 1992. 

\bibitem{DS} E. Doron  and U. Smilansky, \textit{Semiclassical quantization of chaotic billiards: a scattering theory approach}, Nonlinearity \textbf{5},  1055  (1992).

\bibitem{two_point_sqrt_n} D. El-Baz, J. Marklof, and I. Vinogradov, \textit{The two-point correlation function of the fractional parts of $\sqrt{n}$ is Poisson}, Proc.   AMS  \textbf{143}, 1  (2013). 


\bibitem{sqrt_n} N.  D.  Elkies  and  C.  T.  McMullen, \textit{Gaps  in $\sqrt{n}$ mod 1  and  ergodic  theory}, Duke Math. J. \textbf{123}, 95 (2004).

    
\bibitem{noble} B. Noble, \textit{Methods based on the Wiener-Hopf technique}, Chelsea Publishing Company, New York, N. Y. (1988).



\end{thebibliography}
\end{document}